\documentclass{svjour3}                     
\usepackage{color}
\usepackage{amssymb,amsfonts}
\usepackage{graphicx}
\journalname{--}
\begin{document}
\large
\title{\textcolor[rgb]{0.00,0.00,1.00}{A New Formulation of Electrodynamics}
}


\author{\textcolor[rgb]{1.00,0.00,0.00}{Arbab I. Arbab}  \emph{and}   \textcolor[rgb]{1.00,0.00,0.00}{Faisal. A. Yassein}
}


\institute{Arbab I. Arbab \at
              Department of Physics, Faculty of Science, University of Khartoum, P.O. Box 321, Khartoum
11115, Sudan \\
              Tel.: +249-122-174858\\
              Fax: +249-183780539\\
              \email{arbab.ibrahim@gmail.com, aiarbab@uofk.edu}           
           \and
           F. A. Yassein \at
             Department of Physics, Faculty of Science, Alneelain  University, P.O. Box 12702, Khartoum, Sudan\\
              \email{f.a.yassein@gmail.com}
}

\date{Received: 12 September 2009 / Accepted: }

\maketitle

\begin{abstract}
A new formulation of  electromagnetism based on linear differential commutator brackets is developed. Maxwell equations are derived, using these commutator brackets, from the vector potential $\vec{A}$, the scalar potential $\varphi$ and the Lorentz gauge connecting them. With the same formalism, the continuity equation is written in terms of these new differential commutator brackets.
\end{abstract}

\section{Introduction}
\baselineskip=20pt
 Maxwell equations are first order differential equations in space and time. They are compatible with Lorentz transformation which guarantees its applicability to any inertial frame.  A symmetric space-time formulation of any theory will generally guarantees the universality of the theory. With this motivation, we adopt a differential commutator bracket involving first order space and time derivative operators to formulate the Maxwell equations and quantum mechanics. This is in addition to our recent quaternionic formulation of physical laws, where we have shown that many physical equations are found to emerge from a unified view of physical variables [1]. In such a  formulation, we have  found that Maxwell equations emerges from a single equation. Maxwell equations were originally written in terms of quaternions. They were initially written in twenty equations [a]. However, later on Maxwell equations are then written in terms of vector in the way that we are familiar today. In our present formulation, Maxwell equations are described by a set of two wave equations representing the evolution of the electric and magnetic fields. This is instead of having four equations. We aim in this paper to write down (derive) the physical equations by vanishing  differential commutator brackets. We know that second order partial derivatives commute for space-space variables. We don't assume here this property is a priori for space and time. To guarantee this, we eliminate the time derivative of a quantity that is acted by a space ($\nabla$) derivative followed by a time derivative, and vice-versa. In expanding the differential commutator bracket, we don't commute time and space derivative, but rather eliminate the time derivative by the space derivative, and vice versa. This differential commutator bracket may enlighten us to quantize these physical quantities.
By employing the differential commutator brackets of the vector $\vec{A}$ and scalar potential $\varphi$, we have derived Maxwell equations without invoking any  a priori physical law. Maxwell arrives at his theory of electromagnetism by combing  the Gauss, Faraday and Ampere laws. For mathematical consistency, he modified Ampere's law. He then came with the known Maxwell  equations.
\section{Relativistic prelude}
From Lorentz transformations one obtain,
\begin{equation}\label{1}
  x'=\gamma(x-vt)\,,\qquad y'=y\,,\qquad z'=z \,,\qquad t'=\gamma(t-\frac{v}{c^2}\,x) \,.
\end{equation}
We see that the commutator bracket
\begin{equation}\label{1}
\left[\triangle t\,, \triangle x\right]=\left[ \triangle t'\,, \triangle x'\right]\,.
\end{equation}
where we have taken into account in the order of multiplication of the space and time differences, ($\triangle x\,, \triangle t$). This shows that the commutator is Lorentz invariant. This is a new invariant quantity in relativity. We, however, already knew that the square interval is Lorentz invariant, i.e., $(\triangle S)^2 =(\triangle S')^2$ [2]. It follows from Eq.(1) that the differential commutator bracket $\left[\frac{\partial}{\partial t}\,, \vec{\nabla}\right]=0\,$ is Lorentz invariant too, i.e., $\left[\frac{\partial}{\partial t}\,, \vec{\nabla}\right]=\left[\frac{\partial}{\partial t'}\,, \vec{\nabla}'\right]$.
We know that the spatial second order derivatives of a function, $f=f(x,y)$, is commutative, i.e., $\frac{\partial^2f}{\partial x\partial y}=\frac{\partial^2f}{\partial y\partial x}$. We wonder if the commutation of space and time derivatives are equally valid for all physical quantities.   Motivated by this hypothesis, we propose the following differential commutator brackets to formulate the physical laws. In particular, we apply these differential commutator brackets, in this work to derive  the continuity equation, Maxwell equations.
\section{Differential commutators algebra}
Define the three linear differential commutator brackets as follows:
\begin{equation}\label{1}
    \left[\frac{\partial}{\partial t}\,, \vec{\nabla}\right]=0\,,\qquad    \left[\frac{\partial}{\partial t}\,, \vec{\nabla}\cdot\right]=0\,,
\qquad  \left[\frac{\partial}{\partial t}\,, \vec{\nabla}\times\right]=0\,.
\end{equation}
Equation (3) is correct since partial derivatives commute, i.e., $\frac{\partial^2}{\partial t\partial x}\varphi=\frac{\partial^2}{\partial x\partial t}\varphi$.
For a scalar $\psi$ and a vector $\vec{G}$, one defines the three brackets as follows:\footnote{See the Appendix for other identities.}
\begin{equation}\label{1}
    \left[\frac{\partial}{\partial t}\,, \vec{\nabla}\right]\psi=\frac{\partial}{\partial t}\left(\vec{\nabla}\psi\right)-\vec{\nabla}\left(\frac{\partial \psi}{\partial t}\right)\,,
\end{equation}
\begin{equation}\label{1}
    \left[\frac{\partial}{\partial t}\,, \vec{\nabla}\cdot\right]\vec{G}=\frac{\partial}{\partial t}\left(\vec{\nabla}\cdot\vec{G}\right)-\vec{\nabla}\cdot\left(\frac{\partial \vec{G}}{\partial t}\right)\,,
\end{equation}
and
\begin{equation}\label{1}
  \left[\frac{\partial}{\partial t}\,, \vec{\nabla}\times\right]\vec{G}=\frac{\partial}{\partial t}\left(\vec{\nabla}\times\vec{G}\right)-\vec{\nabla}\times\left(\frac{\partial \vec{G}}{\partial t}\right)\,.
\end{equation}
It follows that
\begin{equation}\label{1}
    \left[\frac{\partial}{\partial t}\,, \vec{\nabla}\cdot\right](\psi\vec{G})= \psi\{\left[\frac{\partial}{\partial t}\,, \vec{\nabla}\cdot\right]\vec{G}\}+\{ \left[\frac{\partial}{\partial t}\,, \vec{\nabla}\right]\psi\}\cdot\vec{G}\,,
\end{equation}

\begin{equation}\label{1}
    \left[\frac{\partial}{\partial t}\,, \vec{\nabla}\times\right](\psi\vec{G})= \psi\{\left[\frac{\partial}{\partial t}\,, \vec{\nabla}\times\right]\vec{G}\}+\{\left[\frac{\partial}{\partial t}\,, \vec{\nabla}\right]\psi\}\times\vec{G}\,,
\end{equation}
\begin{equation}\label{1}
    \left[\frac{\partial}{\partial t}\,, \vec{\nabla}\cdot\right](\vec{G}\times\vec{F})= \vec{F}\cdot\{\left[\frac{\partial}{\partial t}\,, \vec{\nabla}\times\right]\vec{G}\}-\vec{G}\cdot \{\left[\frac{\partial}{\partial t}\,, \vec{\nabla}\times\right]\vec{F}\}\,,
\end{equation}
for any vector $\vec{F}$.
The differential commutator brackets above satisfy the distribution rule
\begin{equation}\label{1}
    \left[\hat{A}\hat{B}\,\,,\hat{C}\right]= \hat{A} \left[\hat{B}\,\,,\hat{C}\right]+ \left[\hat{A}\,\,,\hat{C}\right]\hat{B}\,,
\end{equation}
where
$\hat{A}\,, \hat{B}\,,\hat{C}$ are $\vec{\nabla}\,, \frac{\partial}{\partial t}$. 
It is evident that the differential commutator brackets  identities follow the same ordinary vector identities.
We call the three differential commutator brackets in  Eq.(3) the grad-commutator bracket, the dot-commutator bracket and the cross-commutator bracket respectively. The prime idea here is to replace the time derivative of a quantity by the space derivative $\vec{\nabla}$ of another quantity, and vice-versa, so that the time derivative of a quantity is followed by a time derivative with which it commutes. We assume here that space and time derivatives don't commute. With this minimal assumption, we have shown here that all physical laws are determined by  vanishing differential commutator bracket.
\section{The continuity equation}
Using quaternionic algebra [3], we have recently found that generalized continuity equations can be written as
\begin{equation}
\vec{\nabla}\cdot \vec{J}+\frac{\partial \rho}{\partial
t}=0\,,
\end{equation}
\begin{equation}
\vec{\nabla}(\rho\,c^2)+\frac{\partial
\vec{J}}{\partial t}=0\,,
\end{equation}
and
\begin{equation}
\vec{\nabla}\times
\vec{J}=0\,.
\end{equation}
 Now consider the dot-commutator of $\rho\vec{J}$
 \begin{equation}\label{1}
    \left[\frac{\partial}{\partial t}\,, \vec{\nabla}\cdot\right](\rho\vec{J})=\frac{\partial}{\partial t}\left(\vec{\nabla}\cdot (\rho\vec{J})\right)-\vec{\nabla}\cdot\left(\frac{\partial (\rho\vec{J})}{\partial t}\right)=0\,.
\end{equation}
Using Eqs.(11) - (13), one obtains
\begin{equation}\label{1}
    \left[\frac{\partial}{\partial t}\,, \vec{\nabla}\cdot\right](\rho\vec{J})=c^2\rho\left(\frac{1}{c^2}\frac{\partial^2\rho}{\partial t^2}-\nabla^2\rho\right)+\left(\frac{1}{c^2}\frac{\partial \vec{J}}{\partial t^2}-\nabla^2\vec{J}\right)\cdot\vec{J}=0\,.
\end{equation}
For arbitrary $\rho$ and $\vec{J}$, Eq.(15) yields the two wave equations
\begin{equation}\label{1}
 \frac{1}{c^2}\frac{\partial^2\rho}{\partial t^2}-\nabla^2\rho=0\,,
\end{equation}
and
\begin{equation}\label{1}\frac{1}{c^2}\frac{\partial \vec{J}}{\partial t^2}-\nabla^2\vec{J}=0\,.
\end{equation}
Equations (16) and (17) show that the charge and current density satisfy a wave equation travelling at speed of light in vacuum. It is remarkable to know that these two equations are already obtained in [3]. Hence, the current-charge density wave equations are equivalent to
\begin{equation}\label{1}
    \left[\frac{\partial}{\partial t}\,, \vec{\nabla}\cdot\right](\rho\vec{J})=0\,.
\end{equation}
\section{Maxwell's equations}
Maxwell's equations are first order differential equations in space and time of the electromagnetic field, viz.,
\begin{equation}\label{2}
\vec{\nabla}\cdot \vec{E}=\frac{\rho}{\varepsilon_{0}}\,,
\end{equation}
\begin{equation}\label{2}
\vec{\nabla}\times \vec{B}=\mu_{0}\vec{J}+\frac{1}{c^2}\frac{\partial \vec{E}}{\partial t},
\end{equation}
\begin{equation}
\vec{\nabla}\times\vec{E}=-\frac{\partial \vec{B}}{\partial t}\,,
\end{equation}
\begin{equation}\label{2}
\vec{\nabla}\cdot \vec{B}=0\,.
\end{equation}
These equations show that charge ($\rho$) and current ($\vec{J}$) densities are the sources of the electromagnetic field.
Differentiating Eqs.(20) and using Eq.(21), one obtains
\begin{equation}
\frac{1}{c^2}\frac{\partial^2\vec{E}}{\partial
t^2}-\nabla^2\vec{E}=-\mu_0\left(\vec{\nabla}(\rho\,c^2) +\frac{\partial
\vec{J}}{\partial t}\right)\,.
\end{equation}
Similarly, differentiating Eq.(21) and using Eq.(20), one obtains
\begin{equation}
\frac{1}{c^2}\frac{\partial^2\vec{B}}{\partial
t^2}-\nabla^2\vec{B}=\mu_0\left(\vec{\nabla}\times \vec{J}\right)\,.
\end{equation}
These two equations state that the electromagnetic field propagates with speed of light in two cases:\\
(i) charge and current free medium (vacuum), i.e., $\rho=0, \vec{J}=0$, or \\
(ii) if the two equations
\begin{equation}
\vec{\nabla}(\rho\,c^2) +\frac{\partial
\vec{J}}{\partial t}=0\,,
\end{equation}
and
\begin{equation}
\left(\vec{\nabla}\times \vec{J}\right)=0\,,
\end{equation}
besides the familiar continuity equation in Eq.(11)
\begin{equation}
\vec{\nabla}\cdot\vec{J}+\frac{\partial \rho}{\partial t}=0\,,
\end{equation}
are satisfied.
Equation (23) and (24) resemble Einstein's general relativity equation where space-times geometry is induced by the distribution of matter present. We see here that the electromagnetic field is produced by any charge and current densities distribution (in space and time).
Now define the electromagnetic vector $\vec{F}$ as
\begin{equation}
\vec{F}=\vec{B}-\frac{i}{c}\vec{E}\,
\end{equation}
Adding Eqs.(25) and Eqs.(26) according to Eq.(28), one obtains
\begin{equation}
\frac{1}{c^2}\frac{\partial^2}{\partial
t^2}(\vec{B}-\frac{i}{c}\vec{E})-\nabla^2(\vec{B}-\frac{i}{c}\vec{E})=\mu_0\left[\frac{i}{c}\left(\vec{\nabla}(\rho\,c^2) +\frac{\partial
\vec{J}}{\partial t}\right)+\vec{\nabla}\times \vec{J}\right]\,.
\end{equation}
Applying Eqs.(25), (26) (see [3]) in Eq.(29) yields
\begin{equation}
\frac{1}{c^2}\frac{\partial^2\vec{F}}{\partial
t^2}-\nabla^2\vec{F}=0\,\,,\qquad \Box^2\vec{F}=0\,.
\end{equation}
This is a wave equation propagating with speed of light in vacuum ($c$). Hence, Maxwell wave equations can be written as a pure single wave equation of an electromagnetic  sourceless complex vector field $\vec{F}$.
We call Eqs.(25) - (27) the generalized continuity equations. We have  recently obtained these  generalized continuity equations by adopting quaternionic formalism for fluid mechanics [3]. It is challenging to check wether any real fluid satisfies these equations or not.
We have recently shown that Schrodinger, Dirac and Klein - Gordon and  diffusion equations are compatible with these generalized continuity equations [3].
Using Eqs.(19) and (20), the electric field dot-commutator bracket yields
\begin{equation}\label{1}
    \left[\frac{\partial}{\partial t}\,, \vec{\nabla}\cdot\right]\vec{E}=\frac{\partial}{\partial t}\left(\vec{\nabla}\cdot\vec{E}\right)-\vec{\nabla}\cdot\left(\frac{\partial \vec{E}}{\partial t}\right)=\frac{\partial\rho}{\partial t}+\vec{\nabla}\cdot\vec{J}=0\,.
\end{equation}
This is the familiar continuity equation. Hence, the continuity equation in the commutator bracket form can be written as
\begin{equation}\label{1}
    \left[\frac{\partial}{\partial t}\,, \vec{\nabla}\cdot\right]\vec{E}=0\,.
\end{equation}
Similar, using Eqs.(21) and (22), the magnetic field dot-commutator bracket yields
\begin{equation}\label{1}
    \left[\frac{\partial}{\partial t}\,, \vec{\nabla}\cdot\right]\vec{B}=\frac{\partial}{\partial t}\left(\vec{\nabla}\cdot\vec{B}\right)-\vec{\nabla}\cdot\left(\frac{\partial \vec{B}}{\partial t}\right)=0\,.
\end{equation}
The electric field cross-commutator bracket gives
\begin{equation}\label{1}
    \left[\frac{\partial}{\partial t}\,, \vec{\nabla}\times\right]\vec{E}=\frac{\partial}{\partial t}\left(\vec{\nabla}\times\vec{E}\right)-\vec{\nabla}\times\left(\frac{\partial \vec{E}}{\partial t}\right)=0\,.
\end{equation}
Using Eqs.(20) and (21), this yields
\begin{equation}\label{1}
    \left[\frac{\partial}{\partial t}\,, \vec{\nabla}\times\right]\vec{E}= \frac{1}{c^2}\frac{\partial^2\vec{B}}{\partial t^2}-\nabla^2\vec{B}- \mu_0\left(\vec{\nabla}\times\vec{J}\right)=0\,.
\end{equation}
This equation is nothing but Eq.(24) above.
Similarly, the magnetic field  cross-commutator bracket gives
\begin{equation}\label{1}
    \left[\frac{\partial}{\partial t}\,, \vec{\nabla}\times\right]\vec{B}=\frac{\partial}{\partial t}\left(\vec{\nabla}\times\vec{B}\right)-\vec{\nabla}\times\left(\frac{\partial \vec{B}}{\partial t}\right)=0\ .
\end{equation}
Using Eqs.(20) and (21) this yields,
\begin{equation}\label{1}
\left[\frac{\partial}{\partial t}\,, \vec{\nabla}\times\right]\vec{B}= \frac{1}{c^2}\frac{\partial^2\vec{E}}{\partial t^2}-\nabla^2\vec{E}+ \mu_0\left(\vec{\nabla}(\rho\, c^2)+\frac{\partial \vec{J}}{\partial t}\right)=0\,.
\end{equation}
This equation is nothing but Eq.(23) above. Hence, Eqs.(35) and (37), i.e.,
\begin{equation}\label{1}
    \left[\frac{\partial}{\partial t}\,, \vec{\nabla}\times\right]\vec{E}= 0\,,\qquad
\left[\frac{\partial}{\partial t}\,, \vec{\nabla}\times\right]\vec{B}= 0\,.
\end{equation}
represent the combined Maxwell equations. In terms of the vector $\vec{F}$ defined in Eq.(33), the wave equation in Eq.(35) can be written as
\begin{equation}\label{1}
    \left[\frac{\partial}{\partial t}\,, \vec{\nabla}\times\right]\vec{F}= 0\,,
\end{equation}
which is also evident from Eq.(28).
\section{Derivation of Maxwell equations from the vector  and  scalar potentials, $\vec{A}, \varphi$ }

The electric and magnetic fields are defined by the vector $\vec{A}$ and the scalar potential $\varphi$ as follows
\begin{equation}
\vec{E}=-\vec{\nabla}\varphi-\frac{\partial \vec{A}}{\partial t}\,\,,\qquad \vec{B}=\vec{\nabla}\times\vec{A}\,.
\end{equation}
These are related by the Lorentz gauge as
\begin{equation}
\vec{\nabla}\cdot\vec{A}+\frac{1}{c^2}\frac{\partial \varphi}{\partial t}=0\,.
\end{equation}
Comparing this equation with Eq.(11) reveals that the continuity equation is nothing but a gauge condition. This means that a new current density $\vec{J}'$ can be found so that the equation is still intact. We have recently explored such a possibility and showed that it is true [3]. With this motivation the physicality of the gauge $\vec{A}$ exhibited by Aharonov–Bohm effect is tantamount to that of the current density $\vec{J}$[5].
The grad-commutator bracket of the scalar potential $\varphi$
\begin{equation}\label{1}
    \left[\frac{\partial}{\partial t}\,, \vec{\nabla}\right]\varphi=\frac{\partial}{\partial t}\left(\vec{\nabla}\varphi\right)-\vec{\nabla}\left(\frac{\partial \varphi}{\partial t}\right)= 0\,.
\end{equation}
Using Eqs.(40) and (41), one obtains
\begin{equation}\label{1}
    \left[\frac{\partial}{\partial t}\,, \vec{\nabla}\right]\varphi=\frac{1}{c^2}\frac{\partial^2\vec{A}}{\partial t^2}-\nabla^2\vec{A}-\mu_0\vec{J}=0\,.
\end{equation}
This yields the wave equation of the vector field $\vec{A}$ as
\begin{equation}\label{1}
\frac{1}{c^2}\frac{\partial^2\vec{A}}{\partial t^2}-\nabla^2\vec{A}=\mu_0\vec{J}\,.
\end{equation}
Similarly, the dot-commutator bracket of the vector $\vec{A}$
\begin{equation}\label{1}
    \left[\frac{\partial}{\partial t}\,, \vec{\nabla}\cdot\right]\vec{A}=\frac{\partial}{\partial t}\left(\vec{\nabla}\cdot \vec{A}\right)-\vec{\nabla}\cdot\left(\frac{\partial \vec{A}}{\partial t}\right)=0\,.
\end{equation}
Using Eqs.(40) and (41), one obtains
\begin{equation}\label{1}
    \left[\frac{\partial}{\partial t}\,, \vec{\nabla}\cdot\right]\vec{A}=\frac{1}{c^2}\frac{\partial^2\varphi}{\partial t^2}-\nabla^2\varphi-\frac{\rho}{\varepsilon_0}=0\,.
\end{equation}
This yields the wave equation of $\varphi$
\begin{equation}\label{1}
\frac{1}{c^2}\frac{\partial^2\varphi}{\partial t^2}-\nabla^2\varphi=\frac{\rho}{\varepsilon_0}\,.
\end{equation}
The cross-commutator bracket of the scalar potential $\varphi$
\begin{equation}\label{1}
    \left[\frac{\partial}{\partial t}\,, \vec{\nabla}\times\right]\vec{A}=\frac{\partial}{\partial t}\left(\vec{\nabla}\times\vec{A}\right)-\vec{\nabla}\times\left(\frac{\partial \vec{A}}{\partial t}\right)= 0\,.
\end{equation}
Using Eqs.(40), one finds
\begin{equation}\label{1}
    \left[\frac{\partial}{\partial t}\,, \vec{\nabla}\times\right]\vec{A}=\frac{\partial \vec{B}}{\partial t}+\vec{\nabla}\times\vec{E}= 0\,.
\end{equation}
This yields the Faraday's equation,
\begin{equation}\label{1}
\vec{\nabla}\times\vec{E}= -\frac{\partial \vec{B}}{\partial t}\,.
\end{equation}
It is interesting to arrive at this result by using the definition in Eq.(40) only.
Now consider the dot-commutator bracket of $\varphi\vec{A}$
\begin{equation}\label{1}
    \left[\frac{\partial}{\partial t}\,, \vec{\nabla}\cdot\right](\varphi\vec{A}) =\frac{\partial}{\partial t}\left(\vec{\nabla}\cdot(\varphi\vec{A})\right)-\vec{\nabla}\cdot\left(\frac{\partial }{\partial t}(\varphi\vec{A})\right)= 0\,.
\end{equation}
Using Eqs.(40), (41) and the vector identities
\begin{equation}\label{1}
\vec{\nabla}\cdot(\varphi\vec{G})=(\vec{\nabla}\varphi)\cdot\vec{G}+\varphi(\vec{\nabla}\cdot\vec{G})\,,\qquad
\vec{\nabla}\times(\vec{\nabla}\times\vec{G})=\vec{\nabla}(\vec{\nabla}\cdot\vec{G})-\nabla^2\vec{G}\,,
\end{equation}
Eq.(51) yields
\begin{equation}\label{1}
\varphi\left(\vec{\nabla}\cdot\vec{E}-\frac{\rho}{\varepsilon_0}\right)-c^2\left(\vec{\nabla}\times\vec{B}-\frac{1}{c^2}\frac{\partial \vec{E}}{\partial t}-\mu_0\vec{J}\right)\cdot\vec{A}=0\,.
\end{equation}
For arbitrary $\varphi$ and $\vec{A}$, Eq.(53) yields the two equations
\begin{equation}\label{1}
\vec{\nabla}\cdot\vec{E}=\frac{\rho}{\varepsilon_0}\,,
\end{equation}
and
\begin{equation}\label{1}
\vec{\nabla}\times\vec{B}=\mu_0\vec{J}+\frac{1}{c^2}\frac{\partial \vec{E}}{\partial t}\,.
\end{equation}
Equations (54) and (55) are the Gauss and Ampere equations.

Similarly, the cross-commutator bracket of $\varphi\vec{A}$
\begin{equation}\label{1}
    \left[\frac{\partial}{\partial t}\,, \vec{\nabla}\times\right](\varphi\vec{A}) =\frac{\partial}{\partial t}\left(\vec{\nabla}\times(\varphi\vec{A})\right)-\vec{\nabla}\times\left(\frac{\partial }{\partial t}(\varphi\vec{A})\right)= 0\,.
\end{equation}
Using Eq.(40), (41) and the vector identity
\begin{equation}\label{1}
\vec{\nabla}\times(\varphi\vec{G})=(\vec{\nabla}\varphi)\times\vec{G}+\varphi(\vec{\nabla}\times\vec{G})\,,
\end{equation}
 Eq.(56) yields
\begin{equation}\label{1}
\varphi\left(\vec{\nabla}\times\vec{E}+\frac{\partial \vec{B}}{\partial t}\right)-c^2\left(\vec{\nabla}\times\vec{B}-\frac{1}{c^2}\frac{\partial \vec{E}}{\partial t}-\mu_0\vec{J}\right)\times\vec{A}=0\,.
\end{equation}
For arbitrary $\varphi$ and $\vec{A}$, Eq.(58) yields the two equations
\begin{equation}\label{1}
\vec{\nabla}\times\vec{E}=-\frac{\partial \vec{B}}{\partial t}\,,
\end{equation}
and
\begin{equation}\label{1}
\vec{\nabla}\times\vec{B}=\mu_0\vec{J}+\frac{1}{c^2}\frac{\partial \vec{E}}{\partial t}\,.
\end{equation}
 Once again, Eqs.(59) and (60) are the Faraday and Ampere equations, respectively.
Hence, the four Maxwell equations are completed. To sum up, Maxwell equations are the commutator brackets
\begin{equation}\label{1}
 \left[\frac{\partial}{\partial t}\,, \vec{\nabla}\cdot\right](\varphi\vec{A})= 0\,,\qquad  \left[\frac{\partial}{\partial t}\,, \vec{\nabla}\times\right](\varphi\vec{A})=0\,.
\end{equation}

\section{Energy conservation equation}

In electromagnetism, the energy conservation equation  for electromagnetic field is written as
\begin{equation}\label{1}
 \frac{\partial u}{\partial t}+\vec{\nabla}\cdot\vec{S}=-\vec{J}\cdot\vec{E}\,,
\end{equation}
where
\begin{equation}\label{1}
 u=\frac{1}{2}\,\varepsilon_0E^2+\frac{1}{2\mu_0}B^2\,\,,\qquad \vec{S}=\frac{\vec{E}\times\vec{B}}{\mu_0}\,.
\end{equation}
The energy conservation equation of the electromagnetic field can be easily obtain using the following vector identity
\begin{equation}\label{1}
\vec{\nabla}\cdot(\vec{F}\times\vec{G})=\vec{G}\cdot(\vec{\nabla}\times\vec{F})-\vec{F}\cdot(\vec{\nabla}\times\vec{G})\,.
\end{equation}
Let now $\vec{E}=\vec{F}\,\,,\,\,\vec{G}=\vec{B}$, so that  Eq.(64) becomes
\begin{equation}\label{1}
\vec{\nabla}\cdot(\vec{E}\times\vec{B})=\vec{B}\cdot(\vec{\nabla}\times\vec{E})-\vec{E}\cdot(\vec{\nabla}\times\vec{B})\,.
\end{equation}
Employing Eqs.(20), (21) and (63), Eq.(65)  yields
\begin{equation}\label{1}
 \frac{\partial u}{\partial t}+\vec{\nabla}\cdot\vec{S}=-\vec{J}\cdot\vec{E}\,,
\end{equation}
which is the familiar energy conservation equation of the electromagnetic field [5].

\section{Concluding Remarks}

By introducing  three vanishing linear differential commutator brackets for scalar and vector fields, $\varphi$ and $\vec{A}$ and the Lorentz gauge connecting them, we have derived the Maxwell's equations and the continuity equation without resort to any other physical equation.  Using different vector identities, we have found that no any independent equation can be generated from the three differential commutators brackets.

\section*{Acknowledgements}
Equations (14) and (15) are in the form of coupled wave equations known
as inhomogeneous Helmholtz equations. We see that the current density ~ J
enters into these equations in a relatively complicated way, and for this reason
these equations and are not readily soluble in general.
\vspace*{-2pt}
This work is supported by the university of Khartoum research fund. We are grateful for this support.

\section*{References}
$[1]$ Arbab, A. I., and Satti, Z., \emph{Progress in Physics}, \textbf{2}, 8 (2009).\\
$[2]$ Rindler, W., \emph{Introduction to Special Relativity}, Oxford University Press, USA (1991).\\
$[3]$ Arbab, A. I., and Widatallah, H. M., \emph{The generalized continuity equations}, unpublished (2009).\\
$[4]$ Jackson, J. D., \emph{Classical Electrodynamics}, Wiley, New York, 2nd edition, (1975).\\
$[5]$  Aharonov, Y., and Bohm, D.,  \emph{Significance of electromagnetic potentials in quantum theory}, \emph{Phys. Rev.} \textbf{115}, 485 (1959).\\

\newpage
\section*{Appendix}
$$
    \left[\frac{\partial}{\partial t}\,\,,\vec{\nabla}\times\right](\vec{F}\times\vec{G})=\vec{F}\times\left(\frac{\partial}{\partial t}(\vec{\nabla}\times\vec{G})-\vec{\nabla}\times\frac{\partial\vec{G}}{\partial t}\right)+\left(\vec{\nabla}\times\frac{\partial\vec{F}}{\partial t}-\frac{\partial}{\partial t}(\vec{\nabla}\times\vec{F})\right)\times\vec{G}\qquad (A1)
$$
$$
    \left[\frac{\partial}{\partial t}\,\,,\vec{\nabla}\right](\vec{F}\cdot\vec{G})=\vec{F}\left(\frac{\partial}{\partial t}(\vec{\nabla}\cdot\vec{G})-\vec{\nabla}\cdot\frac{\partial\vec{G}}{\partial t}\right)+\left(\vec{\nabla}\cdot\frac{\partial\vec{F}}{\partial t}-\frac{\partial}{\partial t}(\vec{\nabla}\cdot\vec{F})\right)\vec{G}\qquad\qquad\qquad\qquad (A2)
$$
\end{document}